\documentclass[conference]{IEEEtran}
\usepackage{graphicx} 
\usepackage[margin=0.7in,footskip=0.25in]{geometry}
\usepackage[parfill]{parskip}
\usepackage{amsmath}
\usepackage{cleveref}
\usepackage{caption}
\usepackage{algorithm}
\usepackage{algpseudocode}
\usepackage{subcaption}

\title{SuperUROP: An FPGA-Based Spatial Accelerator for Sparse Matrix Operations}

\author{Rishab Parthasarathy}
\date{April 25, 2025}
\begin{document}
\maketitle

\begin{abstract}
Solving sparse systems of linear equations is a fundamental problem in the field of numerical methods, with applications spanning from circuit design to urban planning. These problems can have millions of constraints, such as when laying out transistors on a circuit, or trying to optimize traffic light timings, making fast sparse solvers extremely important. However, existing state-of-the-art software-level solutions for solving sparse linear systems, termed iterative solvers, are extremely inefficient on current hardware. This inefficiency can be attributed to two key reasons: (1) poor short-term data reuse, which causes frequent, irregular memory accesses, and (2) complex data dependencies, which limit parallelism. Hence, in this paper, we present an FPGA implementation of the existing Azul accelerator~\cite{azul}, an SRAM-only hardware accelerator that achieves both high memory bandwidth utilization and arithmetic intensity. Azul features a grid of tiles, each of which is composed of a processing element (PE) and a small independent SRAM memory, which are all connected over a network on chip (NoC). We implement Azul on FPGA using simple RISC-V CPU cores connected to a memory hierarchy of different FPGA memory modules. We utilize custom RISC-V ISA augmentations to implement a task-based programming model for the various PEs, allowing communication over the NoC. Finally, we design simple distributed test cases so that we can functionally verify the FPGA implementation, verifying equivalent performance to an architectural simulation of the Azul framework.
\end{abstract}

\section{Introduction}
Solving sparse systems of linear equations lies at the core of many scientific computing problems, from circuit layout to urban planning~\cite{aguerre2017urban, davis2012circuits}. From a mathematical perspective, these systems are represented as $Ax = b$, where $A$ is a matrix that represents a set of constraints, $b$ is a vector that represents a set of desired outputs, and $x$ is a vector that represents an unknown set of system parameters. Specifically, these systems are termed as sparse when the majority of values in $A$ are zero, with the percentage of non-zero values potentially being as low as  $0.01\%$~\cite{alrescha}. 

The current widely-adopted solution to this problem involves systems we term \emph{iterative solvers}. Iterative solvers work by making an initial guess for the unknown $x$ and refining their guess until reaching an arbitrary precision away from the correct answer~\cite{tiwari2017pcg}. These solvers are especially useful and widely used with large systems, as for large $A$, directly solving for $x = A^{-1}b$ can be too expensive, or even computationally intractable~\cite{tiwari2017pcg}.

\begin{figure}[h!]
    \centering
    \includegraphics[width=0.9\linewidth]{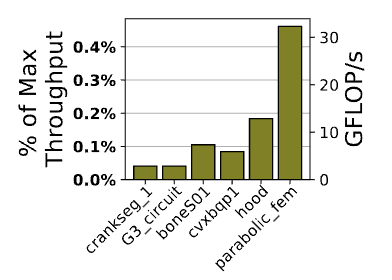}
    \caption{GPUs have extremely low performance when running iterative solvers on sparse matrices, reaching less than 0.5\% of peak throughput. Figure from the Azul paper~\cite{azul}.}
    \label{fig:gpu_gflops_sparse}
\end{figure}
However, despite the mathematical elegance of iterative solvers, they are very inefficient on current hardware, such as CPUs and GPUs, because of frequent memory accesses and hard-to-exploit parallelism. For example, using a standard iterative solver method termed PCG, ~\Cref{fig:gpu_gflops_sparse} demonstrates that on a benchmark of characteristic sparse matrices, GPUs achieve less than 0.5\% of their peak possible throughput, far below a level that would make these algorithms practical to run at scale for the scientific computing community. 

Sparse iterative solvers have two key reasons for they struggle on current hardware: (1) limited data reuse and (2) irregular parallelism. First, since each matrix $A$ in the linear system is large, it takes many megabytes in memory. Each iteration of the iterative solver iterates through the entire matrix and uses each value exactly once, which means that the matrix is evicted from the cache and directly accessed from main memory, leading to slow memory accesses. Second, because the matrices are sparse, the dependencies between data when solving each iteration are irregular, which means that traditional methods like thread-level parallelism and Single Instruction, Multiple Data (SIMD) are ineffective.

Current accelerators like ALRESCHA and DPUv2 have attempted to fill this gap in sparse iterative solving by addressing the second issue of irregular parallelism. Specifically, ALRESCHA and DPUv2 utilize a small number of compute and logic units, which each implement dataflow execution, meaning that the data parallelism is built-into the processor cores themselves. These advancements in dataflow engines enable both ALRESCHA and DPUv2 to saturate the memory bandwidth of their processors, achieving speedup over existing architectures like GPUs. However, these architectures are still bottlenecked by limited data reuse, capping overall performance at the bandwidth of main memory used~\cite{alrescha, dpuv2}.

Hence, recent works have explored the key insight that we can erase the main memory bottleneck of sparse iterative solvers by exploiting \textbf{\emph{inter-iteration reuse}}. These works use \emph{inter-iteration reuse} to refer to the fact that iterative solvers access the same matrix $A$ over thousands of iterations, meaning that keeping the entire matrix in memory would ensure reuse. However, storing the entire matrix in a large cache is insufficient. To ensure fast memory access times for the whole matrix, the memory locations must also be spatially located nearby the processor executing the computation.

A specific accelerator that utilizes \textbf{\emph{inter-iteration reuse}} is Azul, a \emph{spatial accelerator} with large local static memory (SRAMs) placed next to small processing elements (PEs) arranged in a grid. Each of the processing elements is connected to the others via a network-on-chip (NoC), which synchronizes the various PEs through a message-passing communication framework. Each of the PEs owns its own chunk of the matrix $A$, ensuring fast memory accesses and data reuse. On top of that, to emulate the dataflow engines proposed by ALRESCHA and DPUv2, we propose a \emph{task-based programming model}. Instead of directly synchronizing PEs, which would fail to exploit the limited parallelism of the iterative solving problem, tasks are separately on each PE, only stalling execution when necessary to pass messages between different execution units. In this way, Azul implements fine-grained instruction-level parallelism (ILP), matching the dataflow engines of the existing specialized accelerators~\cite{alrescha, dpuv2}.

Hence, in this paper, we investigate the feasibility of implementing Azul on an FPGA, designing each individual component in RTL. We implement a simple FPGA-based version of Azul, where each PE consists of a simple 32-bit RISC-V core, synthesized onto Xilinx FPGAs. We find that when functionally verified, the FPGA implementation matches an architectural simulation in C. This demonstrates that extracting inter-iteration reuse presents a path forward for efficient evaluation of sparse matrices in scientific computing applications, especially when compared to competing architectures like GPUs. 

\section{Background}
In this section, we address related work that this paper builds on, starting from the algorithmic perspective with sparse linear solvers, and later moving to accelerator architecture.

\subsection{Sparse Linear Solvers}
Sparse linear solvers decompose the task of solving linear systems into a combination of two steps that can iteratively repeated until the system is solved to arbitrary precision: (1) Sparse Matrix-Vector multiplication (SpMV) and (2) Sparse Triangular Solves (SpTRSV). 

In SpMV, a matrix $M$ is multiplied by a vector $v$, and the output $y$ is calculated with the relation that
\begin{equation}
    y_i = \sum_{j \in nz(M_i)} M_{ij} v_j
\end{equation}
where $nz$ is a function that returns all nonzero indices in a certain vector. This operation is extremely memory bound, as each matrix element is only used once after being read, but has data parallelism because the calculations of the $y_i$ are functionally independent \cite{spatula, alrescha, dpuv2}.

\begin{figure}[thpb]
    \centering
    \includegraphics[width=0.9\linewidth]{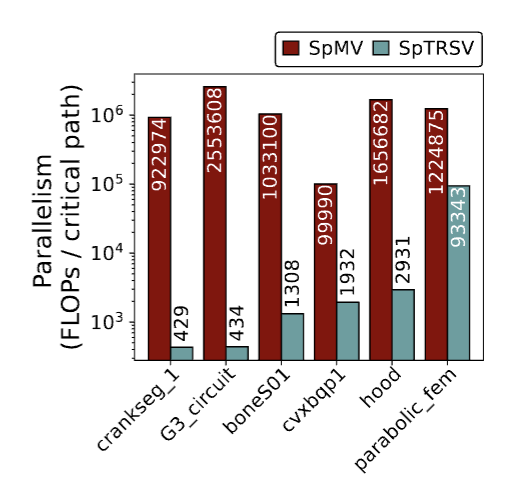}
    \caption{Measured on a set of benchmark matrices, while not having parallelism as high as SpMV, SpTRSV still has high parallelism available from data overall. Figure from Azul paper~\cite{azul}.}
    \label{fig:sptrsv-parallel}
\end{figure}

In SpTRSV, a lower triangular matrix $L$ and a vector $b$ are compared, where the aim is to find a vector $x$ such that $Lx = b$. Unlike SpMV though, this operation is not clearly parallel. For example, $x_0 = \frac{b_0}{L_{00}}$ is easily calculated, but we have $L_{10}x_0 + L_{11}x_1 = b_1$, meaning that the computation for $x_1$ is dependent upon $x_0$. However, even with this dependence, as in \Cref{fig:sptrsv-parallel}, theoretical analysis shows that SpTRSV still has a significant amount of data parallelism, even though it may not be immediately obvious how to leverage it \cite{spatula, alrescha, dpuv2}.

\subsection{Prior Accelerators}
A number of other accelerators have provided functionality for hardware-level implementations of sparse linear solvers, but they all still struggle with the limited data dependencies described in the previous section. Current GPU implementations, despite having high compute throughput, are severely bottlenecked by main memory. Because each iteration of the iterative solver processes the full matrix, the small caches on the GPU cannot capture reuse across iterations, resulting in slow accesses to main memory that throttle overall bandwidth.

Existing accelerators have aimed to solve this problem in two ways. The first is by using targeted dataflow execution. One such example is ALRESCHA, which aims to maximize the parallelism possible within one iteration of both SpMV and SpTRSV, making each iteration of the solver as fast as possible. However, this results in a similar bottleneck to GPUs when scaled up, resulting in limited memory bandwidth and high energy usage from frequent memory accesses \cite{alrescha}. DPU-v2 targets the same problem as ALRESCHA from a different angle, instead processing dataflows at the compiler level rather than the hardware level. By processing at a compiler level, DPU-v2 schedules tasks onto processing elements in a way that maximizes reuse in the cache, but still cannot overcome the limitations of memory bandwidth when iterating through a large matrix \cite{dpuv2}.

Alternatively, some modern accelerators instead follow a distributed SRAM architecture, which means that each small processor owns its own large chunk of very fast memory, where a piece of the array can be stored. These architectures are relatively recent, pioneered by the Cerebras WSE-2, which demonstrated that large distributed SRAM arrays could achieve higher throughput than comparable GPUs on highly parallel tasks like matrix multiplication and deep learning \cite{cerebras}. Recent works have also applied distributed SRAM architectures to linear solvers, demonstrating that these architectures also achieve speedups on non-sparse linear solves across iterations, where large amounts of parallelism are available, but memory bandwidth is still a significant limitation \cite{dalorex}. 

However, it remains to be seen in the literature whether distributed SRAM architectures can be successful on the sparse linear solving tasks. Hence, in this paper, we explore whether combining the key insights from these two directions of research: (1) dataflow execution and (2) distributed memory, can produce an accelerator that addresses both inter-iteration data reuse and can exploit large amounts of irregular parallelism is feasible on FPGA.

\subsection{Azul}

\begin{figure*}[thpb]
    \centering
    \includegraphics[width=1\linewidth]{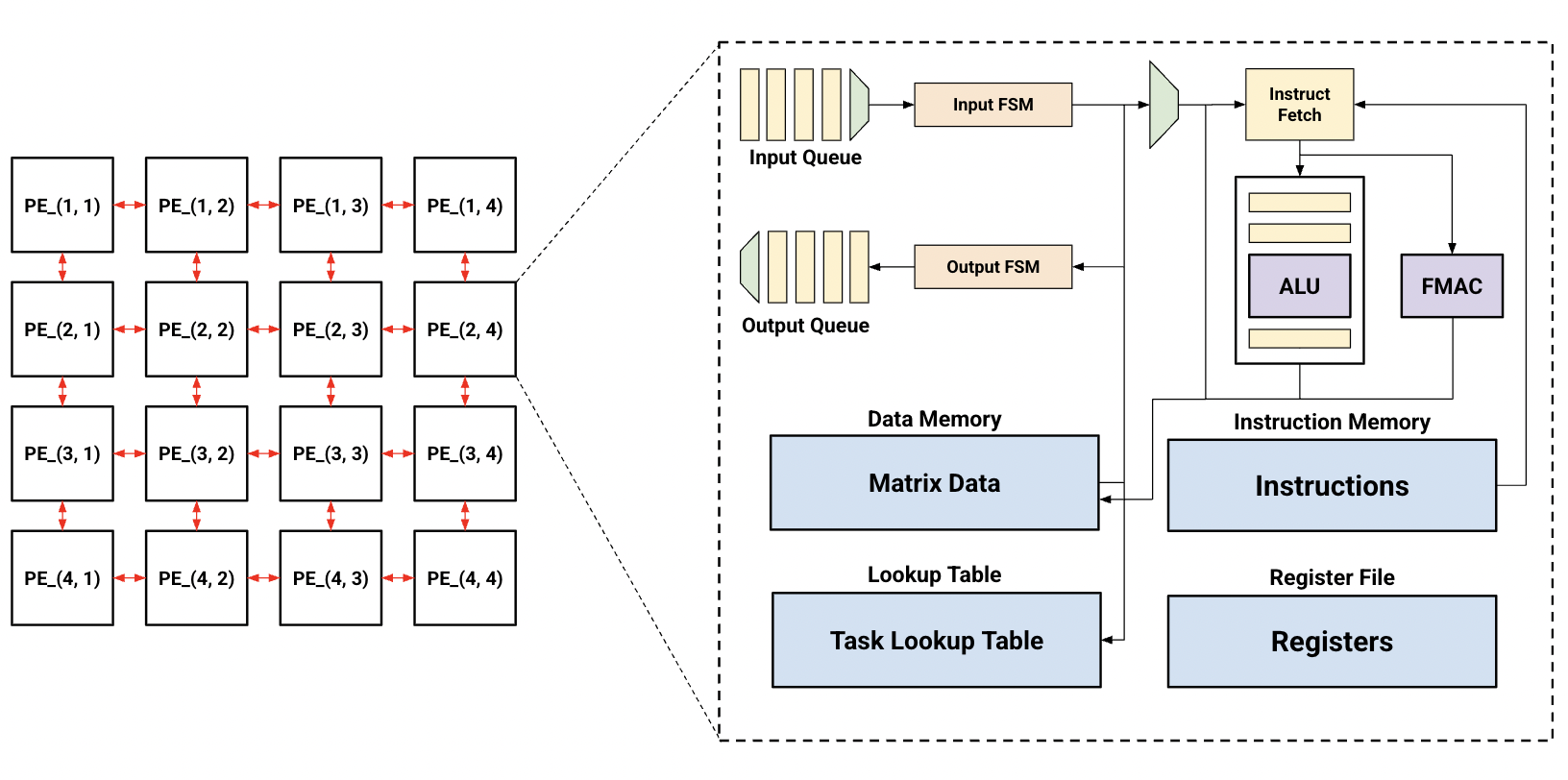}
    \caption{This figure presents an overview of the Azul architecture. The processing elements are arranged in a grid, with each PE containing network input and output queues, which are linked to an integer processor with a floating point multiplier. The memory is divided into subsystems for data memory, instruction memory, and lookup table memory.}
    \label{fig:azul-overview}
\end{figure*}
Specifically, we build on the work done in the design of the Azul sparse matrix accelerator, which is presented in \Cref{fig:azul-overview}. Azul is a distributed SRAM-only spatial accelerator targeted at the tasks of sparse linear solvers, where the SRAM on each tile of the accelerator serves as a host for program memory, data memory, and network input/output queues. Each tile thus consists of a small RISC-V core as a processing element (PE), network queues for connecting to other processing elements in the grid, and the aforementioned memory scratchpad, which serves as extremely rapid access memory \cite{fpga-spmv, fpga-matmul, azul}.

Azul both handles the memory bottleneck of SpMV and SpTRSV while still exploiting the limited parallelism of SpTRSV, a task that is difficult for current GPUs to perform \cite{alrescha, spatula, dalorex, azul}. Specifically, one of the key advancements of Azul is that the sparse data within the matrix $A$ in $Ax = b$ can be partitioned into blocks, which are assigned to individual tiles of the accelerator. Once partitioned to blocks, since the matrix itself does not change -- only the proposed $x$ changes over multiple iterations of sparse linear solvers -- the memory access time to perform SpMV and SpTRSV operations is greatly reduced, as there is no need to pull values from a slow backing DRAM or higher level cache, as on a standard GPU. We call this new form of parallelism we exploit in Azul, \emph{\textbf{inter-iteration parallelism}}, as the processing of sparse matrices into blocks is a one-time expense that can be offloaded to a compiler or precomputation framework \cite{mppa, pim-graph, dalorex, azul}.

The other key advancement of Azul is adopting a \emph{\textbf{a task based instruction-passing framework}}. Instead of complex control flow frameworks, where multiple processor elements are parallelized by having the same instruction operate on different data (often called SIMD), we instead make Azul follow a task-based abstraction. In this so-called task-based abstraction, each PE executes a task (or set of instructions) independently of all other PEs, blocking as needed to send data or receive data from other PEs across the network. This fine-grained task-based control limits the generalization capabilities of Azul, but allows for exploiting parallelism where fine-grained control is needed, such as in SpTRSV, where the rapid task-switching enabled by this programming model allows Azul to quickly switch between various data paths without blocking due to data dependencies.

\section{Methods}

In this section, we explore the design decisions taken in designing Azul on FPGA.

\subsection{PE Architecture}

The processing element is implemented using a simple pipelined RISC-V processor, as the ALU must only need to implement simple operations like addition and memory access. These simple operations can fit in one clock cycle of a standard 1 ns clock period, which is why we choose a simpler processor over a more advanced one. Outside of the ALU, we attach a floating point multiplier for more complicated operations on SpMV and SpTRSV.

The RISC-V processor consists of five stages. The first is an instruction fetch stage that retrieves the instruction from the SRAM scratchpad, which passes into a decode stage. This decode stage then provides an instruction to the ALU, which finally sends the output to writeback to be written to register file or main memory. These stages are represented by the registers after the instruction fetch in \Cref{fig:azul-overview}.

\subsubsection{Memory Architecture}
The memory architecture is what allows this lightweight processor architecture to still exploit significant parallelism across multiple iterations of data. Unlike standard GPUs and accelerators, where there are multiple levels of memory between the data and the computation, such as HBM, L1 caches, and scratchpads, each PE instead has an extremely large SRAM tile allocated to it, which allows for single-cycle random access to every element within it. As such, by storing the entire block of the sparse matrix $M$ allocated to the block in SRAM, the memory architecture enables single-cycle reads to every value within the matrix. The SRAM tile also contains all the instructions required for evaluating the tasks required of the PE. 

\begin{figure*}[thpb]
    \centering
    \includegraphics[width=0.9\linewidth]{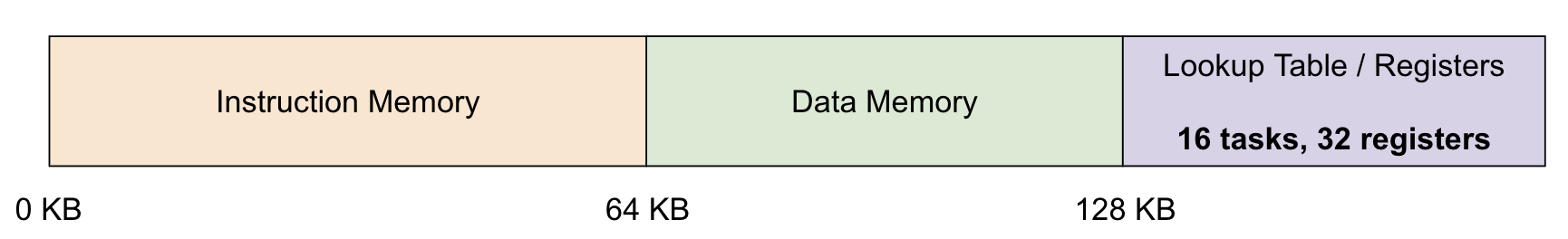}
    \caption{This figure describes the memory hierarchy of each SRAM tile. The bottom 64 kB are reserved for instruction memory, the next 64 kB for data memory, and higher hanging SRAM for a lookup table and register file.}
    \label{fig:memory-hierarchy}
\end{figure*}

The specific partitioning of this SRAM tile can be found in \Cref{fig:memory-hierarchy}. The bottom 64 kilobytes are allocated for instruction memory, while the next 64 kilobytes are allocated for data memory. These values are chosen so that both instruction memory and data memory can be accessed using 16-bit addresses. On top of the instruction and data memory lies a 32-wide 32-bit register file, which is used to store results of intermediate computations. Finally, the lookup table is a 16-wide table that is used to initialize the task-level abstraction of the instruction memory. While this limits the number of tasks and type of tasks that is placed, the lookup table maps from task identifier to starting address in instruction memory, which allows for the tasks to be easily registered as a partition of the SRAM memory. These memory sizes were chosen so that the entire SRAM tile of each PE could fit in the native BRAM and URAM of the FPGA instead of mapping to the larger, but significantly slower DRAM. We make this choice as the few-cycle latency of BRAM and URAM more closely mimics the real world performance of SRAM.

\subsection{Instruction Set Augmentation}
\label{sec:isa}

\begin{figure*}[thpb]
    \centering
    \includegraphics[width=0.9\linewidth]{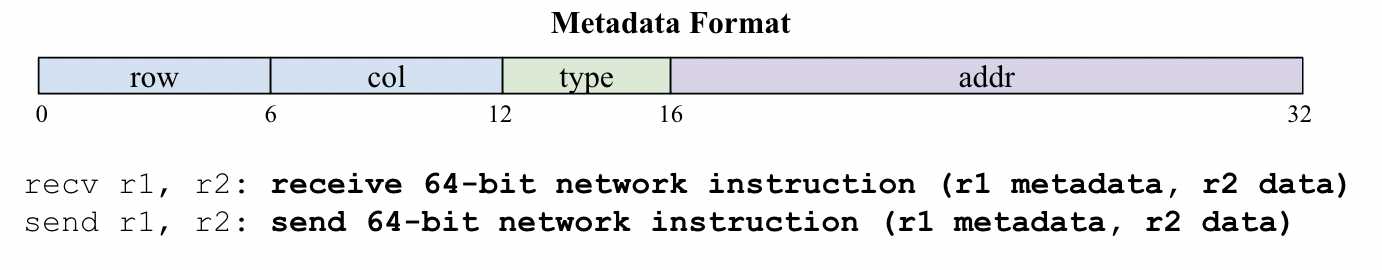}
    \caption{The metadata for the networking augmentation contains the row/column, the task type, and the address to write to, which are used by the send and recv message augmentations.}
    \label{fig:metadata}
\end{figure*}
To support task-based programming, where communication across the network is the only means of synchronization instead of having synchronized instruction streams like SIMD, we augment the standard 32-bit RISC-V ISA with two instructions, as described in \Cref{fig:metadata}. The $send$ and $recv$ instructions formally codify receiving and sending instructions from the networking FIFOs, as this remains the only source of communication between PEs in a fully distributed SRAM-only scheme. While this communication scheme results in network bottlenecks on extremely large data and tasks without parallelism, choosing this approach allows for higher utilization of the inherent parallelism of the sparse linear problem in this case.

Specifically, as shown in \Cref{fig:metadata}, the $send$ and $recv$ instructions encode 64-bit networking values as a metadata register and a data register. The data register contains the data to be written/read from the register file, and the metadata register contains the row/column of the target PE in the PE grid as 6-bit fields, a 4-bit task type, and a 16-bit address. This limits the number of rows and columns to 64x64, but 4096 tiles is a sufficiently large amount that we consider this size to be sufficient. The task type multiplexes between a write to instruction memory, data memory, or lookup table, with a special fourth task type being the start of the task indicated by the address to write in the lookup table. These semantics allow a global controller to write different tasks and trigger different tasks on individual PEs, which then synchronize each other via the network when executing tasks.

\subsubsection{Programming Model}
\begin{algorithm*}
\caption{Azul Task Execution on a Processing Element (PE)}
\label{alg:azul_task_execution}
\begin{algorithmic}[1]
    \Statex \textbf{Phase 1: Network Reading}
    \Loop
        \State Read message $m$ from Network Input Queue

        \State Process network message $m$ (e.g., write to memory based on $m$.type and $m$.addr) 
    \EndLoop
    \Statex
    \Statex \textbf{Phase 2: Task Execution Cycle}
    \State Set Program Counter $pc \leftarrow 0$\\
    \Loop \Comment{Idle loop}
        \State Wait for Network Input Queue to be non-empty
        \State Read message $m$ from Network Input Queue
        \If {$m$.type is START\_TASK}
            \State Look up task start address: $task\_pc \leftarrow \text{LookupTable}[m.\text{addr}]$
            \State Set Program Counter $pc \leftarrow task\_pc$ 
            \State Execute task instructions starting from $pc$ 
            \State Set Program Counter $pc \leftarrow 0$ \Comment{Task returns, PE idles}\\
        \Else
            \State Process other network message $m$ (e.g., handle incoming data during idle)
        \EndIf
    \EndLoop
\end{algorithmic}
\end{algorithm*}

As described in ~\Cref{alg:azul_task_execution}, the programming of Azul has two phases: (1) network reading and (2) task execution. In network reading, each PE reads messages from its network until all data memory, instruction memory, and lookup memory has been sufficiently written for execution. In task execution, the processor idles at until receiving a network message that indicates the start of a task. Then, the device jumps to the $pc$ stored within the lookup table for the task, meaning that task execution ends. When a task completes, the Azul dataflow engine returns to the idle state.

Each task is implemented as a function in a standard language like C++. These functions directly exploit parallelism by exposing the $send$ and $recv$ instructions to the programmer using assembly code injection, enabling the user to control communication, along with controlling the PEs that certain data/tasks are sent to. Azul does not provide any guarantees about ensuring deadlock security or about the correctness of the data tiling performed--this is enforced on the side of the programmer.

\section{Evaluation}
In this section, we discuss the experimental methodology and evaluation for our FPGA implementation of Azul.

\subsection{FPGA Implementation}

To implement this system onto FPGAs, we implemented the PEs in Minispec to process the full instruction set detailed in \Cref{sec:isa}, which was synthesized at a 200 MHz clock period on a cluster of Xilinx FPGAs. To synthesize the different memory domains, we mapped the different sections of memory to different chunks of the FPGA memory. Lookup table memory is mapped to the high performance lookup LUTRAM, while the register file is mapped to the RAM32M 32-bit register file Xilinx primitive. Finally, data and instruction memory is mapped to the single-cycle URAM, and the network FIFOs to the similarly single-cycle BRAM. These decisions are made as the URAM is the largest memory and thus contains the bulk of memory used by the processor. All FPGA compilation and place-and-route is performed using Vivado.

\subsection{Testing Configuration}

For testing configuration, we validate all FPGA implementations against a cycle-accurate simulation of a 16x16 tiled architecture, running at a 2GHz clock frequency. The network topology is a 2D torus, with the same memory sizes as the FPGA implementation (64 KB per PE)~\cite{azul}. 

\subsection{Test Cases}

We first test our implementation against a set of simple distributed task-based test cases. These test cases are designed by interleaving multiplication and add compute tasks with the \texttt{send} and \texttt{recv} ISA augmentations, testing whether simple dataflow patterns in SpMV and SpTRSV cause deadlock and align properly to the expectations of the simulated architecture.

\begin{figure*}[htbp] 
    \centering 

        \includegraphics[width=\linewidth]{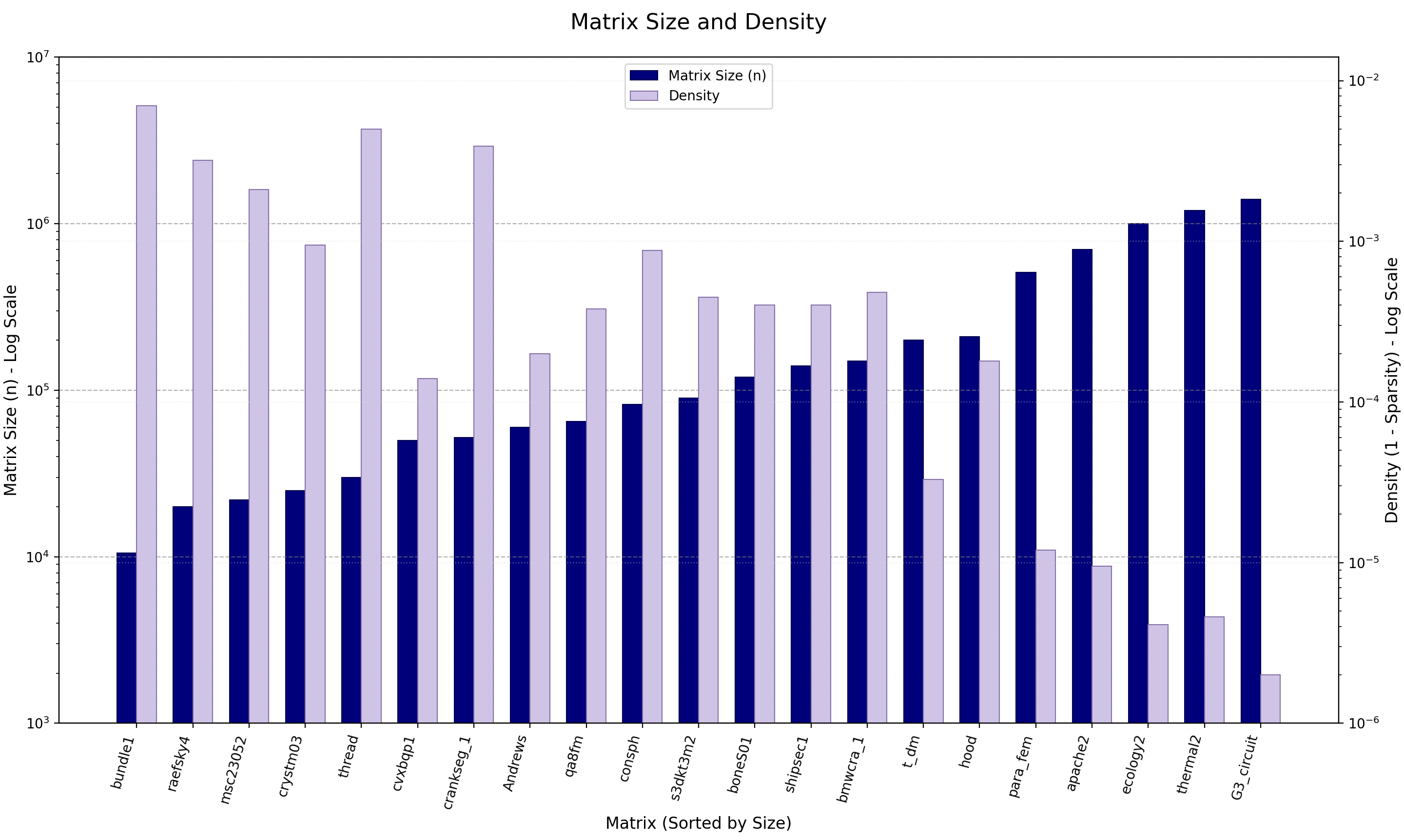}
        \caption{A sampling of matrix sizes and densities from the SuiteSparse benchmark.}
        \label{fig:results-size1}


\end{figure*}

Outside fo simple test cases, we evaluate on matrices from the SuiteSparse benchmark, a small subset of which are described in ~\Cref{fig:results-size1}.

\subsection{Evaluation Results}

Here, as discussed in the Methods section, the key difference from the simulation is that the chip-level SRAMs in Azul are implemented with a combination of large BRAMs and URAMs on FPGA. In addition, an FPGA must implement the network-on-chip using logic and compute units, resulting in reduced network bandwidth and potential network latency issues. As a result, we validate whether the FPGA variant of Azul remains compute-bound rather than network or memory bottlenecked, enabling efficient use of compute relative to power consumption.

Overall, we find that the FPGA implementation of Azul remains compute-bound even on FPGA, matching expected results calculated through a Python script. Through our functional verification scheme, we verify the performance of both small toy test cases and large test cases from the SuiteSparse benchmark. From the small test cases, we find that the \texttt{send} and \texttt{recv} instructions properly implement networking queues and do not cause deadlock or unnecessary network latency. From the larger test cases taken from the SuiteSparse benchmark, we find that we match performance of simulation when the memory latency is matched between simulation and real FPGA hardware along with matching a sample Python implementation, confirming numerical validity as well.
\section{Conclusion}

In this paper, we have presented an FPGA implementation of Azul, a spatial accelerator for sparse iterative solvers that combines the paradigms of dataflow execution and distributed SRAM memory~\cite{azul}. We design a spatial grid of RISC-V processing elements integrated with floating point multiplication units, connected via a Network-on-Chip. To model distributed SRAM memory on FPGA, we propose a custom memory structure, partitioning data, instruction, register, and lookup table memory between the available LUTRAM, BRAM, URAM, and XRAM on FPGA to maximive performacne. Finally, in combination with a custom RISC-V instruction set extension and task-based programming model, our FPGA implementation of Azul allows for easy use of dataflow-based computation through the task lookup table implemented on each PE. Comparing to a simulation of the Azul microarchitecture and sample Python testbenches, we functionally verify this FPGA implementation on both small toy test cases and larger test cases from the SuiteSparse benchmark. In general, we have demonstrated that existing consumer hardware like FPGAs can be used to implement Azul, potentially meaning that Azul can be used as a competitor to GPUs for any user with access to FPGAs. In the future, with continued research to increase the arithmetic intensity and clock frequency of FPGA implementations of such spatial accelerators, this work shows that sparse matrix problems may become as simple as a drag-and-drop solution (where the user compiles the desired RTL code onto their FPGA and is able to execute the task faster than the fastest GPUs).
\section{Acknowledgements}
The author would like to thank Prof. Daniel Sanchez and Dr. Axel Feldmann for their mentorship and for introducing the author to the Azul architecture. This author would also like to thank Citadel and the SuperUROP program at MIT for sponsoring this work.
\bibliographystyle{unsrt} 
\bibliography{refs} 

\end{document}